\documentclass{aastex}

\shorttitle{Vector Gravitational Waves}

\shortauthors{Latapiat et~al.}

\begin{document}

\title{Wheeler's Gravitomagnetism Revisited I:
       A Purely Lorentz-Compton Approach to Vector Gravitational Waves
       and Trio-Holes}

\author{Latapiat, Roberto A.}
\affil{Rabbi (Tzfat), Dean of Ieshiva kol Iaakov, Caracas, Venezuela}

\author{Miranda, Guillermo F.}
\affil{unigrav@hotmail.com \\
       Escuela de Matem\'atics, Facultad de Ciencias, \\
       Universidad Central de Venezuela,
       Ciudad Universitaria, Caracas, Venezuela}

\author{Pereyra, Nicolas A.}
\affil{pereyrana@utpa.edu \\
       Department of Physics and Geology,
       College of Science and Engineering, \\
       University of Texas - Pan American,
       1201 W. University Drive,
       Edinburg, TX 78541-2999}

\begin{abstract}

A theory of vector gravitational waves is presented on the basis of an
additional gravitational field,
recovering ideas that go back to H. A. Lorentz,
field that A. Wheeler called ``gravitomagnetism''.
The theory is independent of Einstein's General Relativity Theory,
but is consistent with the Einstein-Lorentz framework of Special
Relativity,
so that the resulting equations are considered exact from the start,
and not approximations valid for velocities small compared with the
speed of light.
A simple model of mass formation and presence of angular momentum
started with a new type of singularity,
``trio-holes'',
is exhibited.

\end{abstract}

\section{Introduction}

The main theoretical purpose of the present work is to exhibit
gravitation as a particular form of interaction between matter and
light (and more generally radiation),
within the framework of Special Relativity,
but in a manner wholly independent of Einstein's General Theory.

The mathematical modeling of gravitational phenomena has not reached
yet the state of development provided for electromagnetic phenomena by
the Maxwell-Lorentz system of linear partial differential equations and
its further quantization by Dirac, Feynman and others
\citep[see][]{fey1965,fey1985}.
After Faraday introduced the notion of an electromagnetic field as a
physical entity on its own right,
the classical theory of electromagnetism,
in spite of Coulomb's inverse square distance Law of force,
could be considered free from an action at a distance postulate only
thanks to Maxwell's equations.

In Einstein's own words \citep{ein1950}:
``It was Maxwell who fully comprehended the significance of the field
concept;
he made the fundamental discovery that the laws of electrodynamics
found their natural expression in the differential equations for the
electric and magnetic fields.
These equations implied the existence of waves,
whose properties corresponded to those of light as far as they were
known at that time....
The new insight made it possible to dispense with the hypothesis of
action at a distance,
at least in the realm of electromagnetic phenomena;
the intermediary field now appeared as the only carrier of
electromagnetic interaction between bodies,
and the field's behavior was completely determined by contiguous
processes,
expressed by differential equations....
Because one cannot dispense with the field concept,
it is preferable not to introduce in addition a carrier with
hypothetical properties.''

Lorentz,
did much to extend Maxwell's work into the atomic scale,
introducing the Lorentz force and explained the normal Zeeman effect
for the first time \citep{lor1909}.
According to L. De Broglie \citep{deb1939} a physicist image of
reality needs to change when transporting it from the human or
astronomical scale onto the much smaller atomic scale,
where Quantum Laws cannot be avoided so far.
In this first work we shall not consider the analog of the quantization
procedure applied to electrodynamics,
considered by Dirac and followers,
leaving aside the atomic scale.

There are many reasons to hold on to the group of Lorentz
transformations for as long as doing this yields fruitful results.
Seemingly,
Maxwell's equations are here to stay at least in the human and
astronomical scales,
and as Lorentz discovered early,
they are covariant (for empty space) with respect to Lorentz
transformations.
While it is true,
as Einstein observes,
that ``Maxwell's equations imply the Lorentz group,
but the Lorentz group does not imply Maxwell's equations'',
it is also true that ``... it is to be expected that all equations of
physics are covariant with respect to Lorentz transformations
(special theory of relativity)'' \cite{ein1950}.

Einstein further declares that:
``A consistent field theory requires continuity of all elements of the
theory,
not only in time but also in space,
and in all points of space.
Hence the material particle has no place as a fundamental concept in a
field theory.
Thus even apart from the fact that gravitation is not included,
Maxwell's electrodynamics cannot be considered a complete theory.'' 

We doubt that such a statement,
dating back to 1950,
would have been written after the advent of Laurent Schwartz's theory of
distributions \citep{sch1959},
one of the major mathematical breakthroughs of the 20th century,
together with Lebesgue's measure and integration theory (1901) and
Hilbert and Banach's infinite dimensional normed vector spaces,
among other mathematical advances pertinent to mathematical physics
(of course, pertinency changes with time).
Distribution Theory gave a precise mathematical meaning to the familiar
Dirac's ``delta function $\delta$''
(introduced intuitively in the nineteen-thirties),
defining it as a continuous linear functional acting upon a topological
(non normed) infinite dimensional vector space of well behaved test
functions,
thus allowing for the correct handling of partial distributional
differentiation of all orders,
and also making full sense of $\Delta(1/r) =  - 4 \pi \delta$ as an
equality among distributions (``r'' denoting the distance of a variable
point in $E_3$ to the origin 0).

Schwartz's original work was technically difficult,
involving the theory of topological vector spaces,
but afterwards,
Vladimirov's and Zemanian's expositions
\citep[see][]{vla1984,zem1987} made the
subject accessible to all scientists dealing with linear differential
equations.

In the case of non-homogeneous linear partial differential equations
(PDE)
with constant coefficients,
taken in a distributional sense,
and with a Dirac's delta distribution used as a forcing term,
the mathematically singular fields obtained by solving them,
called ``fundamental solutions'',
can be usefully associated with the physical fields generated by
physical monopoles,
as exemplified by the fundamental solution for the  Laplace's operator.
The distributional derivatives of Dirac's delta adequately symbolize
multipoles,
the delta itself standing for a monopole.
Due to the linearity of the distributional differential equations,
the distributional solution when the forcing term is a multipole is
obtained from the fundamental solution by distributional
differentiation.
Besides,
the fundamental solution can always be obtained by means of the
distributional Fourier Transform when it cannot be obtained by
inspection \citep[see][]{vla1984,zem1987}.

In regions of regularity,
the distributional solution obtained can be identified with an ordinary
differentiable function,
recovering classical solutions.
Just about the only thing you cannot do with distributions is to
multiply them in a manner that would be analogous to the familiar
non-linear terms of classical PDE's and fields,
except in some particular cases not devoid of interest for a physicist,
such as those worked out by two of the present authors in connection
with a uniqueness problem set up by Fritz John for the non-linear
conservation law defined by $u_t + u u_x = $ in the domain $t>0$,
$-\infty < x < \infty$ \citep{mir2007a},
and solved by  means of a ``moving'' delta starting from (0,0),
with an intensity varying with time.

Should a linear field theory for gravitation be amenable to a
reasonable mathematical justification,
and also to experimental verification in either the astronomical scale
or human scale laboratories at the beginning,
and also at the atomic scale after some adequate quantization procedure
had been completed,
then by the distributional handling of the proposed linear PDE's for
adequately defined gravitational fields,
it would have been established a much simpler way of dealing with known
gravitational facts (or allegedly so) and also with unexplained
features related to angular momentum in the astronomical scale.
This is the plan we propose to follow in the present work.
Besides,
the way the theory is reasoned,
makes of it a more natural step towards a unified treatment of gravity
and electromagnetism.

Einstein sought to unify electromagnetism and gravitation by
introducing a curved space-time geometrization of the gravitational
field first (general theory of relativity),
and then trying to geometrize electromagnetism.
Our point of view will be the opposite,
namely,
we will ``electromagnetize'' (so to speak) gravitation.
  
There have been other theories of gravitation presented to the
scientific community with varying evidence of experimental support
\citep[see][]{her1978,all1995b,all1995a,mas2003,sal1998},
and Einstein's 1928 work ``Einheitlishe feldstheorie von Gravitation
und Elektrizitat'',
trying to obtain this unifying goal,
was criticized by Pauli and others,
and Einstein died without success in this area.

In Wheeler's book \citep{whe1990},
it is asserted that ``... mass,
going round and round in a circle,
must produce a new kind of force,
gravitomagnetism.
Oersted discovered the creation of magnetism from moving electric
charges in 1820,
but gravitomagnetism has not yet been discovered,
although it is the target of two great research enterprises
(Gravity Probe B and LAGEOS project)''.

The term ``gravitomagnetism'' coined by Wheeler,
is sometimes called ``Gravitoelectromagnetism'' and abbreviated as GEM
by some recent authors \citep[see][]{mas2003}.
The recent trend is to refer to it only as a set of formal analogies
between Maxwell's field equations and an approximation to the Einstein
field equations for general relativity,
valid only for slowly moving particles.

In this work,
GEM will be given a different interpretation,
not depending upon Einstein's general relativity approach to
gravitation,
and where Wheeler's force,
plus other forces postulated by us,
will be allowed to act not only upon mass particles,
but also upon light quanta and conversely.

We shall motivate this approach with additional historical
considerations.
As early as 1918,
Schr\"odinger \citep{shr1918} made a bright objection to Einstein's
General Relativity handling of the stress-energy tensor,
to which Einstein finally answered as follows:
``... while sympathizing with these concerns,
I am nonetheless convinced that a definition of the energy components of
the gravitational field more appropriate than the one I have given is
not possible'' \citep{ein1918}.  No further comments.

Again,
even as late in Einstein's life as 1950,
he was cautious about his general theory of relativity and unified
field theory,
and writes \citep{ein1950}
``The editors of Scientific American have asked me to write
about my recent work which has just been published.
It is a mathematical investigation concerning the foundations of field
physics ...'', and also ``As for my latest theoretical work,
I do not feel justified in giving a detailed account of it before a
wide group of readers interested in science.
That should be done only with theories which have been adequately
confirmed by experience''.
These are the words of a great scientific mind indeed,
and set an example to be followed,
notwithstanding the fact that a theoretical work may precede its
experimental confirmation,
as it happened with Dirac's theoretical prediction of a positron,
among examples of this fact taken from ``modern physics'' as
distinguished from ``classical physics'' in the sense given to these
terms by L. De Broglie,
the founder of Wave Mechanics \citep{deb1939}.
Besides,
De Broglie emphasizes that modern physics is based on experiment,
``constituting a method which deliberately produces certain given
conditions in order to see what phenomena are caused by these
conditions'' \citep{deb1939}.

More precisely,
De~Broglie (1939)
sets up a four steps procedure to obtain a sound physical theory:
\hfill \break
i)	The experimental method allows certain laws to be established.
\hfill \break
ii)	Theory then interprets these laws by establishing a connection
	with a single principle.
\hfill \break
iii)	Theory uses the principle thus formulated in order to predict
	quantitatively other phenomena.
\hfill \break
iv)	Finally, recourse is had once again to experiment to verify the
	precision of the predictions made by theory.
\hfill \break
We shall adhere to this guiding procedure in relation with our proposed
model of vector gravitational waves in the final section of this
article,
and shall also refer to ``experiments'' already considered in this area.

In his 1937 book,
De Broglie, having begun with a Chapter entitled
``A general survey of present-day physics'',
in Chapter IV, section 4 ``Relativity and Quanta'' writes:
``In its earlier form, the Theory of Relativity was concerned with the
space and time co-ordinates only of observers in an state of uniform
motion in a straight line. Later it was generalized by Einstein himself,
and has provided a description of Gravitation. This generalized
Relativity however,
I shall here leave on one side'' \cite{deb1939}.

In another book,
published the same year in French \cite{deb1952}
De Broglie writes the following words to which we adhere:
``But if the Special Theory of Relativity seems to be sufficiently
confirmed by experiment, we believe it convenient to be a little less
definite in that which refers to the Generalized Theory.
The new phenomena which it predicts are in fact very small,
and even though they may be observed,
one can always ask whether they are really originated by the cause
which Einstein's Theory attributes them or by some other perturbing
cause,
very small and negligeable while reasoning.
Neither the very weak precession of Mercury's perihelion nor the very
weak light rays deviation passing near the sun disk seem to supply an
irrefutable proof of the exactness of the relativistic conception about
gravitation:
these phenomena do exist and have the order of magnitude predicted by
Einstein's Theory,
but his interpretation does not impose at all upon us.''

Besides Einstein's cautious words about his Generalized Theory already
quoted,
we might also mention the following episode concerning Feynman's first
professional talk in February 1941 held at the Princeton Physics
Colloquium.
Pauli objects to Feynman's lecture,
and then Pauli said ``Don't you agree, Professor Einstein?'',
and Einstein's reply to Pauli was
``no,
the theory seemed possible,
perhaps there was a conflict with the theory of gravitation,
but after all the theory of gravitation was not so well
established ....'' \citep{gle1993}.

Recently,
(July 2004)
Stephen Hawking gave a talk on his new theory of quantum gravity at his
department in Cambridge University.
He used a complex mathematical technique called the
``Euclidean path integral,'' first used by Richard Feynman,
it has generally been applied to subatomic particles,
but Hawking applies the idea to black holes \citep{pep2004}.

Having thus reviewed what have been the main proposals for a theory of
Gravitation,
together with the mathematical and physical contexts which likely
played a part in their formulation,
we shall proceed to detail our proposal.

\section{Vibrations and Generalized Vector Waves}

The starting point here is the following obvious general consideration
made by De Broglie \citep{deb1939}:
``whatever magnitude it is that is traveling in the form of waves must
pass through a periodic variation,
the period itself being obviously equal to the time elapsing between
the passing of two consecutive crests.''
These periodic variations result in a vibration at any given point.
Thus we can define a physical wave as a vibration of some physical
magnitude which travels through space-time with a definite velocity.
Wave velocities are usually associated with characteristic values for
hyperbolic systems of PDE's.
Waves in continuous media whose origin is due to an initial disturbance
imparted to the propagating media,
and not to a vibration forced at some points of the media,
also fit with De Broglie's general definition of waves,
since the initial disturbance can be interpreted as an initial energy
provided to a continuously distributed system,
formed by an infinite number of simple oscillators arranged as
contiguous differential elements capable of vibration at each point of
the media.
Besides physical waves,
we have to consider Schr\"odinger's ``probability waves'',
and their origin is altogether different,
so that the following considerations do not apply to them.     
Let us recall two of the most familiar and simple free vibrating
systems,
in order to define a generalized simple vibration system.
\hfill \break
a) The system mass-spring.
If ``x'' denotes the horizontal displacement at time ``t'' of a mass
``m'' attached to a spring with stiffness ``k'',
which moves without friction, and defining $v= dx / dt$,
Newton's third Law of motion yields:

\begin{equation}
m {dv \over dt} + kx = 0
\;\;\;\; .
\end{equation}

\noindent
Multiplication by ``$v$'' followed by time integration yields:

\begin{equation}
{m \over 2} v^2 + {k \over 2} x^2 = constant
\end{equation}

\noindent
b) A closed circuit formed by a coil or solenoid with inductance $L$
and a condenser with capacitance $C$.
Defining ``$q$'' as the electrical charge at a time ``$t$'' stored by
the condenser,
and letting $i = dq / dt$ stand for the electical current density,
circulating in a coil with negligeable ohmic resistance,
Kirchhoff's Law yields:

\begin{equation}
L {di \over dt} + {1 \over C} q = 0
\;\;\;\; .
\end{equation}

\noindent
Multiplication by $i$ followed by time integration yields:

\begin{equation}
{L \over 2} i^2 + {(1 / C) \over 2} q^2 = constant
\end{equation}

We see,
generalizing from these two classical examples,
that the simplest physical systems
(characterized by ``lumped'' parameters),
in order to vibrate,
require that they are able to store energy in two different forms
[kinetic energy of the mass ``m'' and elastic potential energy of the
spring for example a),
magnetic energy associated to the magnetic flux of the coil,
and electric energy associated to the condenser's electric field for
b)].

So,
if we have two scalar magnitudes $u_1$ and $u_2$,
capable of representing two different energies when squared,
and further related by $u_2 = du_1 / dt$,
they will vibrate when the physical law determined by the nature of the
system is of the general form $C_1u_1 + C_2 (du_2 / dt)=0$,
where $C_1$ and $C_2$ are two positive constants describing some lumped
parameters,
and this second differential equation when multiplied by
$du_1 / dt$  and integrated exhibits the
``Conservation of Energy'' Law for this system:
$C_1(u_1)^2+C_2(u_2)^2 =  constant$.

De Broglie (1939) writes:
``The importance of the part that certain physical magnitudes play
result from the fact that there exists conservation theorems for these
magnitudes''.
This is why the fundamental ordinary differential equation for our
generalized system of two first order equations must be of the form
$- du_2 /  dt$  proportional to  $u_1$ 
after defining $u_2=du_1 / dt$.
Since ${d \over dt}(u_1)^2 = 2u_1 {d \over dt}(u_1)=2u_1u_2$,
this way of relating $u_2$ and $u_1$ ensures that when $u_2$ vanishes,
then the absolute value of $u_1$ reaches a maximum in time,
that is to say,
at a time when one form of the system's total energy vanishes,
the other form of energy reaches its maximum value,
so that a total transfer has been made from one form of energy to the
other,
setting the basis for the oscillatory process.
This basis is completed by the second relation between one magnitude
and the time derivative of the other,
namely,
$u_1=-(C_1 / C_2) (d u_2 / dt)$,
which in view of the first one is equivalent to the conservation law:

\begin{equation}
{C_1 \over 2} (u_1)^2 + {C_2 \over 2} (u_2)^2 = constant
\end{equation}

The real positive valued constant of integration appearing in the right
hand side,
is the total energy of the system,
which is acquired at some initial time $t_0$ by imparting to the system
initial values $u_1(t_0)$ and $u_2(t_0)$ for the two scalar magnitudes
$u_1$ and $u_2$.
This in turn is accomplished through the action of some
``forcing'' terms,
manifesting physical ``forces'' acting externally to the system
considered during some time interval ending at $t_0$.
If ``b'' denotes the horizontal force acting per unit mass
(``body force'') along the positive x-axis in our mechanical example,
and if ``$\epsilon$'' denotes the ``electromotive force'' set up in the
coil by some externally induced field,
then the differential equations for a) and b) become respectively:

\begin{equation}	              
\hbox{forced a)} \hskip 20pt  m{dv \over dt} + kx = mb
\hbox{;} \hskip 20pt v = {dx\over dt}
\end{equation}

\begin{equation}	              
\hbox{forced b)} \hskip 20pt  L{di \over dt} + (1/C)q = \epsilon
\hbox{;} \hskip 20pt i = {dq\over dt}
\end{equation}

If the body force acts impulsively,
with a large intensity ``I'' through a very short time interval
centered at $t_0$,
it can be mathematically idealized by means of the distribution
$I \delta(t-t_0)$.
The mass' horizontal displacement $x(t)$ is continuous through the
process,
otherwise the spring and mass would break apart,
but the velocity $v(t)$,
would appear as having a jump discontinuity at $t=0$.  

Taking $t_0=0$,
and recalling that a piecewise continuously differentiable function,
such as $v(t)$,
having a jump discontinuity at time $t=0$,
is a regular distribution,
having a distributional derivative composed of a regular part plus a
singular part \citep[see][]{mir2007a},
but still satisfying the distributional version of the Fundamental
Theorem of classical Calculus:

\begin{equation}
\int\displaylimits_{-h}^h v'(t) \, dt = v(h) - v(-h)
\hskip 20pt
\hbox{ {\it h} positive}
\;\;\;\; ,
\end{equation}

\noindent
we can compute the distributional definite integral from (-h) to h of
both sides of the distributional equation defined by:

\begin{eqnarray}
{dv \over dt} + (k/m) x = I \delta && \longrightarrow
\nonumber
\\
v(h)-v(-h) + (k/m) \int\displaylimits_{-h}^h x(t) \, dt = I
\;\;\;\; , &&
\hbox{ since } \hskip 10pt
\int\displaylimits_{-h}^h  \delta(t) \, dt = 1
\;\;\;\; .
\end{eqnarray}
                           
In the real,
physical impulsive process,
``$h$''
(defined so that $v(t)=0$ for $t \le -h$)
must be taken as a finite,
though small,
positive quantity.
In the idealized impulsive process,
we take limits as $h$ tends to zero,
thus getting $v(O^+)=I$ as expected,
as well as $x(0^+)=x(0^-)=0$ in this process.

Another way of sustaining vibrations in such simple systems is by means
of a forcing vibrating term
(forced vibrations)
of the form $b(t) = M \exp(i \omega t)$,
leading to resonance when $\omega^2 = (k/m)$,
``M'' being the complex amplitude of the externally applied body force
with angular frequency ``$\omega$''.

These rather lengthy and known considerations about simple lumped
parameters physical systems,
whose free vibrations are governed by a system of two ordinary
differential equations (ODE) of the first order of a very specific form
tied to Conservations Laws,
are justified as a natural way,
leading from a discrete physical system to a continuous vector field
distributed in three-dimensional space $\hbox{E}_3$,
by means of the analogy that can be established between the
energy-densities of two vector fields within a small volume element at
a point of $\hbox{E}_3$ and time ``t'',
and the pair of magnitudes which vibrate in a lumped parameter system
placed there at that time.

Before going into the most general system,
let us go back to our electrical circuit example.
We know by Gauss' Theorem,
that the electric displacement vector $\vec{D}$ at a surface point of
the condenser's plate with a charge ``$q$'' uniformly distributed over
the plate's surface with area $A$ is such that
$D^2 = q^2 / A^2$,
and this means that the vector $\vec{D}$,
and therefore the electric field $\vec{E}$,
normal to the plate surface,
is proportional to ``q'' at any given instant of time ``t''.
Now, since $q=-LC (di/dt)$ and the mean magnetic induction
$\vec{B}$ established by
the current intensity ``$i$'' circulating through the coil is such that
$B=(L/A)i$,
we see that in the localized small region of space about a point
$x=(x_1,x_2,x_3)$ where we assume our analog lumped parameter system to
be concentrated,
the electric field $\vec{E}$ is proportional to
$(- d\vec{B} / dt)$ since 
$q = - AC (dB / dt) = DA$,
and in the simplest case of a condenser formed by two plane parallel
plates of area A and separation ``W'',
$D/C = EW /A$  and
$- d\vec{B} / dt = (W /A) \vec{E}$.
 Note that $(W/A)$ has the dimension of $1/length$.

In the same way,
the other equation of the first order system,
which is the definition $i = dq / dt$,
together with $B = (L/A) i$ and $q = DA$ tell us that
$(1/L) \vec{B} = d\vec{D} / dt$,
and since we know that for a solenoid $(B/L)$ has the dimensions of
$H/length$,
we can write that the second,
defining equation of the system can be written symbolically as
$d\vec{D} / dt = (1 / length)\vec{H}$,
and the previous one as
$- d \vec{B} / dt = (1 / length) \vec{E}$.

When we consider a distributed system,
i.e.,
the fields $\vec{E}$ and $\vec{H}$ varying in space as well as time,
it is to be expected that the ordinary time derivatives of
$\vec{D}$ and $\vec{B}$ become first partial derivatives with respect
to time,
and that the right hand sides become some first order partial
differential operator of $\vec{H}$ and $\vec{E}$ evaluated at time
``t'' as functions of the space variables $x_1$, $x_2$, $x_3$, since
only the net contributions of $\vec{H}$ and $\vec{E}$ are to be
accounted for when they vary in small volume elements about point
$x = (x_1,x_2,x_3)$.

The first order ``d'' operator appearing in the generalized Stokes'
Theorem for differential forms over smooth manifolds,
is represented by means of the divergence, rot (curl) or gradient
operators in particular cases.
Now,
the divergence operator takes vectors into scalars and the gradient
takes scalars fields into vector valued fields.
Thus,
when only vector valued fields can be the domain and range of a first
order ``d'' operator,
the only remaining one is the rot operator,
and we would naturally write as a system of first order PDE
(that extends the above ODE first order homogeneous system)
the following vector electromagnetic (VEM) equations:

\begin{eqnarray}
{\partial \vec{D} \over \partial t} &=& \vec{\nabla} \times \vec{H}
\\
\nonumber \\
- {\partial \vec{B} \over \partial t} &=& \vec{\nabla} \times \vec{E}
\end{eqnarray}
 
The above heuristic considerations are not meant to ``prove'' anything
since the second equation of our PDE system is simply the differential
form of Faraday's law:
$e.m.f. = - {d \over dt} (magnetic \ flux)$.
They are presented as a motivation for considering a Maxwellian type of
system as a natural vector wave prototype or canonical form,
at least when only two solenoidal vibrating vector fields are capable
of storing the energy of the system.

On the other hand,
the distributed system stemming from the mass-spring lumped mechanical
vibrating system,
extends to the familiar distributed system known as
``the vibrating string or membrane'',
with a body force,
such as gravity,
now incorporated as a forcing term:
$\rho{\partial^2 \vec{u} \over \partial t^2} =
 T \Delta \vec{u} + \rho \vec{b}$,
 with $\vec{b} = - g \hat{k}$,
$\vec{u}(\vec{x},t) = u(\vec{x},t)\hat{k}$,
$\vec{x} = x_1 \hat{i}$,
or $\vec{x}=x_1 \hat{i} +x_2 \hat{j}$,
and $\Delta$ is the one or two dimensional scalar
Laplacian,
according to whether we are dealing with a string or a membrane.
This is certainly a non homogeneous scalar wave equation for the scalar
 unknown $u(\vec{x},t)$,
but in contrast with the previous electromagnetic system,
now appear linear homogeneous boundary conditions at the finite boundary
of the tensed string or membrane,
becoming the first and simplest Quantum Model in physics 
(``$T$'' stands for the horizontal initial tension of the string or
membrane at rest,
``$\rho$''is the mass per unit length or area at rest),
first considered by Bernoulli and Euler yielding a discrete spectrum of
allowable vibration frequencies associated to ``stationary waves''.
The associated eigenvalue problem and its solution was later generalized
into the Sturm-Liouville Theory,
as was anticipated by Poincar\'e in the early 19$^{th}$ hundreds.
The higher dimensional generalization of the vibrating string equation
is,
for an isotropic homogeneous linearly elastic three dimensional body:

\begin{equation}
\rho {\partial^2 \vec{u} \over \partial t^2} =
\vec{\nabla} \cdot T(\vec{u}) + \rho \vec{b}
\;\;\;\; ,
\end{equation}

\noindent
where $T(\vec{u})$ is the stress tensor linearly associated with the
strain tensor for the displacement $\vec{u}(\vec{x},t)$ at time ``t''
of a body point initially at $\vec{x}$.
The asymptotic distribution of the eigenfrequencies for a body of
general shape is the subject of a classical paper by Weyl (1915).
The elastodynamics vector equation,

\begin{equation}
\mu \Delta \vec{u} +
(\lambda + \mu) \vec{\nabla} \vec{\nabla} \cdot \vec{u}
=
\rho {\partial^2 \vec{u} \over \partial t^2} - \rho \vec{b}
\end{equation}

\noindent
[where $\lambda$ and $\mu$ are Lam\'e's constants,
 \citep{mir1969}],
admits two kinds of waves,
which can be identified with the familiar P-Waves (compressional waves)
and S-Waves (shear waves) of Seismology.
P-Waves are not transversal,
and this,
together with its natural quantization under appropriate linear
homogeneous boundary conditions,
prescribed on a finite boundary manifold leading to Sturm-Liouville
type of problems,
make this distributed mechanical model substantially different from the
distributed Maxwell model arising from the electric circuit model.
This fact may lie at the root of the unsuccessful attempts to explain
light propagation by means of a hypothetical elastic carrier or
``ether'',
whose history has been extensively described by Whittaker (1973).
Classical Maxwell-Lorentz VEM equations alone cannot explain quantized
energy states.
Besides,
the elastic displacement vector field $\vec{u}(\vec{x},t)$ is NOT
solenoidal,
and neither the velocity field
${\partial \over \partial t}\vec{u}(\vec{x},t)$ is solenoidal,
so that Maxwell-Lorentz is not the appropriated canonical form for
vector waves in this case.

If we think of $\vec{H}=\vec{u}_1$ as a fundamental unknown vector
field and the magnetic induction $\vec{B} = \mu_0 \vec{H}$ 
(in free space)
as a secondary magnitude,
and think of $\vec{E} = \vec{u}_2$ as fundamental vector field and the
electric displacement $\vec{D} = \epsilon_0 \vec{E}$ as secondary,
we obtain the first order system:

\begin{eqnarray}
\vec{\nabla} \times \vec{u_1}
&=&
\epsilon_0 {\partial u_2 \over \partial t}
\\
\nonumber \\
\vec{\nabla} \times \vec{u_2}
&=&
- \mu_0 {\partial u_1 \over \partial t}
\;\;\;\; .
\end{eqnarray}

In the same manner as we asked how a vibration could be initiated and
sustained in time,
it is natural to ask what can be the sources for fields
$\vec{u}_2 = \vec{E}$ and $\vec{u}_1 = \vec{H}$.

In the case of the electric field,
again Gauss' Theorem tells us that at points of space where an electric
charge of volume density $\rho$ is non zero,
we must have $\vec{\nabla} \cdot \vec{u}_2 = \rho / \epsilon_0$.
If $\vec{u}_2 = - \vec{\nabla} U$
with
$U = (1 / 4 \pi \epsilon_0 r)$,
then
$\vec{\nabla} \cdot \vec{u}_2 = - \Delta U = \delta / \epsilon_0$,
so that $\vec{u}_2$ is the electric field due to an electric monopole
with unit charge at the origin of coordinates,
since $\int\int\int \delta \, dx_1 dx_2 dx_3 = 1$ for definite
distributional volume integrals extended over closed domains having the
origin as an interior point.

So far,
no magnetic monopoles have been found in nature,
in spite of a prolonged search by Dirac and others,
so that one must write always $\vec{\nabla} \cdot \vec{u}_1 = 0$
everywhere.

Additionally,
due to the differential form of Ampere's Law for the
``magnetomotive force'' \citep{ble1959},
moving charges constituting a vector density current $\vec{j}$,
act as an additional source term for $\vec{\nabla} \times \vec{u}_1$;
besides,
when we have to deal with conducting bodies,
$\vec{j} = \sigma \vec{E}$
(Ohm's Law),
and when it is a matter with a ``convection current'' as Lorentz calls
it \citep{lor1909},
$\vec{j} = \rho \vec{v}$,
where $\vec{v}$ is the velocity of the freely moving charge passing
through point $x=(x_1,x_2,x_3)$ at time ``$t$''.
We shall not use Gaussian units as is usually done in most theoretical
presentations,
and this for several reasons,
one of them being that when you do so,
it veils the relation of the resulting wave speed to other important
constants related to the propagating medium.

We shall take $\epsilon_0 = 8.8547 \times 10^{-12} [C^2 N^{-1} m^{-2}$
and $\mu_0 = 4 \pi \times 10^{-7} [N A^{-2}]$,
where we used the rationalized meter-kilogram-second-Coulomb system
employed by Bleaney \& Bleaney (1959).
Coulomb's Law of force in free space has the form

\begin{equation}
\vec{F} = { 1 \over 4 \pi \epsilon_0} q_1 q_2 {\vec{r} \over r^3}
\;\;\;\; ,
\end{equation}
the force being repulsive when both charges have the same sign.

When dealing with dielectric or magnetic media,
two new constants are added,
$\epsilon$ and $\mu$,
which are multiplicative correcting factors,
called the dielectric constant and the permeability of the medium
respectively,
so that for isotropic media,
$\vec{D} = \epsilon \epsilon_0 \vec{E}$
and
$\vec{B} = \mu \mu_0 \vec{H}$.
Of course,
for free space is $\epsilon = \mu = 1$.

Finally,
the corrected first order non-homogeneous system of PDE's for isotropic
media in the presence of sources $\rho$ and $\vec{j}$ adopt the
following Maxwell-Lorentz Form (ML):

\begin{eqnarray}
\vec{\nabla} \times \vec{u_1} &=&
\epsilon \epsilon_0 {\partial \vec{u}_2 \over \partial t} + \vec{j}
\\
\nonumber \\
\vec{\nabla} \times \vec{u_2} &=&
- \mu \mu_0 {\partial \vec{u}_1 \over \partial t}
\\
\nonumber \\
\vec{\nabla} \cdot \vec{u}_1 &=& 0
\\
\nonumber \\
\vec{\nabla} \cdot \vec{u}_2 &=& {\rho \over \epsilon \epsilon_0}
\end{eqnarray}

Writing the vector Laplacian
$\Delta = \vec{\nabla} \vec{\nabla} \cdot -
 \vec{\nabla}\times\vec{\nabla}\times$,
we obtain as usual the following non homogeneous vector wave equations
(VEMW):

\begin{eqnarray}
\Delta \vec{u_2} &=&
\mu \mu_0 \epsilon \epsilon_0 {\partial^2 \over \partial t^2} \vec{u}_2
+ \vec{\nabla}\left(\rho \over \epsilon \epsilon_0 \right)
+ \mu \mu_0 {\partial \vec{j} \over \partial t}
\\
\nonumber \\
\Delta \vec{u_1} &=&
\mu \mu_0 \epsilon \epsilon_0 {\partial^2 \over \partial t^2} \vec{u}_1
- \vec{\nabla} \times \vec{j}
\end{eqnarray}

These two equations will be generalized to Vector Gravitational Waves
(VGW).
In free space $\vec{j} = \rho = 0$  and $\epsilon = \mu = 1$,
and the speed of the vector waves turns out to be ``c'',
the speed of light.
It is important to note that for us,
``free space'' is not necessarily identical to ``vacuum'',
and it will mean only a region of $\hbox{E}_3$ where the source terms
$\vec{j}$ and $\rho$ vanish identically during the time interval of
observation for the field disturbance propagation.
Other than that,
``free space'' will not be further precised for the time being.

The change in sign in the first order time derivatives of the first
two ML equations is necessary for conservation reasons as we have 
pointed out.
There is an asymmetry between the equations for $\vec{u}_1$ and
$\vec{u}_2$ with respect to the source terms:
there is no ``$\vec{j}$'' term for $\vec{\nabla}\times\vec{u}_2$ in the
second ML equation  and there is no ``$\rho$'' term for
$\vec{\nabla} \cdot \vec{u}_1$ in the third ML equation,
but this asymmetry might not be intrinsic if magnetic monopoles should
appear after all,
and with them, a magnetic current $\vec{j}$ term in the second ML
equation and this should be kept in mind with possible generalizations
of the ML system of equations to other pairs of vector fields.

The ML system must be supplemented with the Lorentz force equation,
expressing the force $\vec{F}$ acting on a charged particle with
charge $q$ moving with velocity $\vec{v}$ in a region of space where
both $\vec{u}_1$ and $\vec{u}_2$ coexist:

\begin{equation}
\vec{F} = q ( \vec{u}_2 + \vec{v} \times \mu \mu_0 \vec{u}_1 )
\end{equation}

It should be noted that the source (or forcing) terms $\rho$ and
$\vec{j}$ in finite regions of space are not the only cause possible
for setting the motion of vector waves according to equations VEMW,
because one can think of sources acting ``at infinity''.
This rather vague way of speaking can be given a more precise meaning.

It is known that analytic complex functions of a complex variable $z$
can model bidimensional electrostatic fields.
It is also known that negative powers in a Laurent series expansion
about a singular point
(which we shall take to be zero for simplicity)
can represent the potential fields
(equipotential curves and the orthogonal field lines of force)
of the various multipoles through its real and imaginary parts,
so that $1/z$ stands for an electric dipole oriented along the real
axis, $1/z^2$ stands for an electric quadrupole, etc.
But, what is relevant to our present observation is the following fact:
the transformation of the complex $z$-plane into the complex $w$-plane
defined by $w=1/z$ carries negative powers of $z$ into positive powers
of $w$,
and also when $z$ tends to zero,
the absolute value of $w$ tends to infinity.

If you look at the equipotential curves of the identity function
$f(z)=z$,
namely $x=constant$
together with its field lines $y=constant$,
with $z=x+iy$, $x$, $y$ real variables, $i$=imaginary unit,
then it is clear that such would be the field resulting from the
physical field established between two parallel plane plates with
charges equal in magnitude but opposite in sign,
with the planes of the plates orthogonal to the $x$-axis,
and getting farther and farther away to $x=-\infty$ and $x=+\infty$
respectively.
This is what one would call ``an electric dipole at infinity'' without
ambiguity whatsoever.
Again,
if you look at $f(z)=z^2$,
and take two pairs of parallel plates with opposite electric charges,
but one orthogonal to the $y$-axis,
while the other pair is orthogonal to the $x$-axis,
and let them recede to infinity in an obvious way,
we would have obtained what one would call
``an electric quadrupole at infinity'',
and so on for the higher order multipoles associated with the various
powers of the variable ``z''.

The gist of the above consideration is that, when we speak of a regular
analytic function, such as $f(z)=z^2$,
which we see free of singularities in the finite plane,
in reality we are dealing with ``a quadrupole at infinity'',
so that regular analytic functions are deceptively ``regular'' in the
sense that they hide their singular origin at infinity.
Whenever dealing with the solution of field equations in unbounded
regions,
mathematicians have faced a non uniqueness situation concerning
boundary value problems such as the classical Dirichlet and Neumann
problems of Potential Theory \citep{mir2007b},
and as physicists also have to cope with time and again,
coming up with some particular way of handling it,
as it happens with Sommerfeld's Radiation Conditions for the Helmholtz
equation in domains extending to infinity.
In three space dimensions we have an analog of a Laurent series
expansion with the Spherical Harmonics Expansions for solutions of
Laplace's equation.
Potential Theory is a common mathematical tool for both electrostatics
and the classical Newtonian Theory of Gravitation,
which,
in the context of the Gravitomagnetic Theory which we shall discuss in
this work,
might aptly be called ``Gravitostatics'',
in spite of having been used to describe planetary motions with great
success,
which can be explained by the small error that ignoring gravitomagnetic
effects would have introduced in those observed phenomena,
but which cannot be ignored in other phenomena to be discussed ahead.

What has to be kept in mind from the above considerations is that
``what happens at infinity'' cannot be safely ignored when dealing with
field equations in large regions of space.
Besides,
the idea of modeling cosmic rays as
``singularities coming from infinity'' cannot be discarded a priori.

As a very simple and illuminating example taken from \citep{mir2007b},
of how to deal in a precise way with ``multipoles at the infinity'',
consider Laplace's equation in polar coordinates to be solved under a
Dirichlet boundary data ``d'' in the unbounded domain exterior to the
unit disk centered at the origin: 

\begin{equation}
r {\partial \over \partial r}
\left( r {\partial V \over \partial r} \right)
+ {\partial^2 V \over \partial \phi^2} = 0
\hskip 10pt
\hbox{for}
\hskip 10pt
r > 1
, \;\;
0 \le \phi \le 2 \pi
\end{equation}

The general solution, obtained by Fourier's method is:

\begin{equation}
V(r,\phi) =
{Q \over 2 \pi} \log r
+\sum\displaylimits_{n=1}^{\infty}
  \left(A_n \cos n \phi + B_n \sin n \phi\right)
  \left(C_n r^n + D_nr^{-n} \right) + C
\end{equation}

\noindent 
with $Q$, $C$, $A_n$, $B_n$, $C_n$, $D_n$, arbitrary real constants.

By taking $D_n = - C_n$ we obtain an infinity of solutions $V^*$
vanishing for $r=1$,
which correspond to a source of charge $Q$ located at the origin
(or at the infinity)
plus a series of multipoles of equal strength but opposite sign located
at the origin and at infinity:
 
\begin{equation}
V^*(r,\phi) =
{Q \over 2 \pi} \log r
+\sum\displaylimits_{n=1}^{\infty}
  \left(a_n \cos n \phi + b_n \sin n \phi\right)
  \left(r^n - r^{-n} \right)
\end{equation}

Since $z^n = r^n (\cos n\phi + i \sin n\phi)$ and
 $z^{-n} = r^{-n} (\cos n\phi - i \sin n\phi)$,
after taking adequate real parts,
the connection of the terms of our series with multipoles at the origin
(represented by powers $z^{-n}$)
and with multipoles at infinity
(represented by powers $z^n$) is apparent.

The conclusion is that in order to obtain uniqueness for the Dirichlet
problem in the space of unbounded harmonic functions,
it is necessary to prescribe the total electric flux $Q$ for the field
solution plus the complex amplitude $(a_n-ib_n)$ of the multipole
$z^n$ for each natural integer ``n''.
Thus,
in order to precise an unbounded solution,
we must know beforehand what are the causes of unboundedness present at
infinity in order that our final solution behaves asymptotically well
for each natural ``$n$''.
If the radiation from ``outer space'' comes only from a physical
monopole but not from multipoles,
then only $Q$ must be prescribed in addition to the usual Dirichlet
boundary data ``$d$'', and we must take
$a_n=b_n=0$
for all natural ``n''.
The final,
unique solution is obtained in the form $V=V^*+V_d$,
where $V_d$ is the harmonic function ``regular at infinity''
(i.e., bounded in two dimensions)
such that

\begin{equation}
V_d(1,\phi) = C +
\sum\displaylimits_{n=1}^{\infty}
\left(A_n \cos n\phi + B_n \sin n\phi\right) = d(\phi)
\;\;\;\; .
\end{equation}

\noindent
Asymptotically,
$V$ behaves like $V^*$
since $V^*$ is unbounded. 
Similar reasonings have been used in \citep{pow1987} to obtain
sufficient conditions for uniqueness in the case of exterior Stokes
flows.

Use was made in that paper,
as well as in \citep{pow1986},
of the singular solution of Stokes' system of equations known as
``rotlet'',
and this solution will be found useful for discussing angular momentum
effects of our present Theory of vector gravitational waves.
Full use was made of the ``Oseenlet'' and its multipoles in
\citep{mir1983},
establishing existence and uniqueness in the case of the Dirichlet
exterior boundary value problem for the homogeneous Oseen system,
 defining ``Oseen hydrodynamic potentials'' as analogues of classical
single and double-layer surface potentials. 

Before ending our way of considering the establishing of classical
Vector Electromagnetic Waves (VEMW) starting from necessary conditions
to set up vibrations in a general way,
we may recall that the magnetic field $\vec{H}$ in free space generated
by a constant current distribution that extends to infinity along a
straight filament,
with filamentary current $I$,
is such that by Biot-Savart's Law,
it is tangential at each point of a circle of radius $r$,
centered at the filament and in plane normal to it,
with a magnitude given by $H = I / 2 \pi r$.
Thus in an idealized experiment where a physical plane normal to a thin
long wire conducting $I$ was to be covered by iron filings,
one would then see the filings move to form circular patterns.
This image is an analogue of the vision we have had in mind in order to
set up our proposed equations,
but with a semi-infinite light ray instead of an infinitely long wire.

Rather than anticipating,
as Wheeler did,
that a mass going round and round in a circle,
must produce a new kind of force,
with which we have no quarrel,
we have taken two additional steps in order to define these new
gravitomagnetic forces.

Firstly,
we have considered a constant steady mass current distribution along an
infinitely long straight filament producing circular lines of the new
force on planes normal to it,
generating angular momentum in an obvious way.
This certainly would be accomplished if one thinks of the Generalized
Gravitational Field as having two ``components'' satisfying the
adequate analogue of Maxwell-Lorentz (ML),
which might be called ``gravitomagnetic''
(Generalized Gravitational $\vec{u}_1$)
and ``gravitoelectric'' (Generalized Gravitational $\vec{u}_2$),
with a minus sign preceding the source terms stemming from the fact
that two positive masses attract in gravity,
while two positive charges repel in electromagnetism as Mashhoon
adequately observes.

Secondly,
and here is where the name ``Lorentz-Compton approach'' comes in,
we allow the gravitomagnetic component to be generated not only by a
mass current density $\vec{j}$,
but also by Light itself in the form of a current of photons incoming,
as a semi-infinite light ray,
into a ``Trio-Hole'' to be defined in section 4.
If we had a monochromatic ray of frequency $\nu$ say,
then each photon would have an energy $h \nu$
($h$ denoting Planck's constant),
and exactly as in the Compton effect,
it could behave as a particle of equivalent mass $h\nu/c^2$
and equivalent momentum $(h\nu/c)\hat{u}$,
where ``$\hat{u}$'' is a unit vector in the direction of the light ray,
or normal to the light wave front if preferred.
Highly energetic light beams as generators of matter and of the
characteristics of the Created Universe,
were considered in the oral Jewish traditions kept by Tzfat (Safed)
learned men,
and named ``kav lines'' by them \citep{lat2007}.
They also considered the ``Big Crunch'',
a phenomenon prior to Gamow's ``Big Bang'',
under the name of ``tzim-tzum''.
Thinkers of the past have often had intuitions about our physical world
that surprise us once and again.
For a more recent historical example we can consider Descartes' 1625
Theory of vortices,
whereby he declared that matter should not be made only of atoms,
but also of a much smaller particle,
which he conceived as a ``tourbillon'' (vortex).

The constants $\epsilon_0$,
$\epsilon$,
$\mu_0$,
$\mu$
appearing in (ML) and (VEMW) will have gravitational analogs
$\epsilon_{0g}$,
$\epsilon_g$,
$\mu_{0g}$,
$\mu_g$.
In free space we take $\epsilon_g=\mu_g=1$,
and $\epsilon_{0g}$ will be related to Newton's constant
$G=6.67259 \times 10^{-11} m^3/Kg \cdot sec^2$
by the obvious relation
$\epsilon_{0g} = 1 / 4 \pi G = 1.19 \times 10^9 Kg \cdot sec^2/m^3$.

Concerning $\mu_{0g}$,
two options appear,
one is to assume the speed of Vector Gravitational Waves described
(VGW) below to be the speed of light ``c'',
and the other option would be to measure $\mu_{0g}$ directly and then
compute the resulting wave speed as
$(1 / \mu_{0g} \epsilon_{0g})^{1/2}$.

Adopting the first option with
$c^2=8.987044666 \times 10^{16} m^2/sec^2$,
we get $\mu_{0g}=9.3301241 \times 10^{-27} [m/Kg]$.

With the previously assigned meanings for $\vec{j}$ and $\rho$,
the proposed system (GEM) for the gravitomagnetic field $\vec{u}_1$
and gravitoelectric field $\vec{u}_2$ is:

\begin{eqnarray}
\vec{\nabla} \times \vec{u}_1
&=&
\epsilon_g \epsilon_{0g} {\partial \vec{u}_2 \over \partial t}
- \vec{j}
\\
\nonumber \\
\vec{\nabla} \times \vec{u}_2
&=&
- \mu_g \mu_{0g} {\partial \vec{u}_1 \over \partial t}
\\
\nonumber \\
\vec{\nabla} \cdot \vec{u}_1 &=& 0
\\
\nonumber \\
\vec{\nabla} \cdot \vec{u}_2
&=&
- {\rho \over \epsilon_g \epsilon_{0g} }
\end{eqnarray}

\noindent
And the Vector Gravitational Waves equations (VGW) are:

\begin{eqnarray}
\Delta \vec{u}_2 &=&
\mu_g \mu_{0g} \epsilon_g \epsilon_{0g}
{\partial^2 \vec{u}_2 \over \partial t^2}
- \vec{\nabla} \left( \rho \over \epsilon_g \epsilon_{0g} \right)
- \mu_g \mu_{0g} {\partial \vec{j} \over \partial t}
\\
\nonumber \\
\Delta \vec{u}_1 &=&
\mu_g \mu_{0g} \epsilon_g \epsilon_{0g}
{\partial^2 \vec{u}_1 \over \partial t^2} + \vec{\nabla} \times \vec{j}
\end{eqnarray}

\noindent
where $\vec{j}$ can be a photonic current vector density and not only a
mass current density as in Mashhoon.

The GEM system is not to be considered an approximation to geometric
field equations like Einstein's,
as Mashhoon views them,
but on the contrary,
GEM is regarded as an exact system replacing wholly Einstein's original
geometric model concerning gravity.
Gravitational waves appear as vector waves,
solutions of exact linear wave equations,
unlike Dirac's gravitational waves \citep{dir1960},
which appear as approximations in the form of a D'Alembertian operator
applied to each small magnitude $h_{\mu\nu}$ resulting from the
deviation of the fundamental tensor $g_{\mu\nu}$ with respect to the
metric tensor associated to a totally flat space-time when the
Gravitation Constant is taken to be so small that the contracted
Curvature Tensor $R_{\mu\nu}$ can be approximated to zero.
After neglecting terms of higher order in $\gamma$,
Dirac obtains from Einstein's fundamental field equations the
following:  

\begin{equation}
\sq h_{\mu\nu} + {\partial V_\mu \over \partial x^\nu}
+ {\partial V_\nu \over \partial x^\mu}
=
16 \pi \gamma \rho_{\mu\nu}
\hbox{, with}
\hskip 10pt
\sq = \Delta - {1 \over c^2} {\partial^2 \over \partial t^2}
\end{equation}

Dirac further chooses a Coordinate System so that the
``harmonic conditions'' $V_\mu = 0$ are satisfied,
so that an inhomogeneous wave equation,
expressed in this particular Coordinate System,
is satisfied for each of the  $h_{\mu\nu}$'s:

\begin{equation}
\sq h_{\mu\nu}
=
16 \pi \gamma \rho_{\mu\nu}
\hbox{, with}
\hskip 10pt
\rho_{\mu\nu} = T_{\mu\nu} - g_{\mu\nu} T_\lambda^\lambda / 2
\end{equation}

Finally,
in domains where no matter is found,
so that the material energy tensor $T_{\mu\nu}$ vanishes,
Dirac gets $\sq h_{\mu\nu}= 0$ as the equation to describe waves
traveling in free space with the speed of Light ``c''.

In this paper, 
Dirac coins the word ``gravitons'' for particles of Gravitation's
energy traveling with the speed of Light ``c'',
and states that the transfer from these Gravitational waves to
Quantum Theory with the accompanying Hamiltonian Formalism is wholly
not complicated (`ganz unkompliziert'). 
For details concerning nomenclature,
we refer to Dirac's standard textbook \citep{dir1975},
where it is explicitly recognized that 
``$t_\mu^\nu$  (the energy and momentum of the gravitational field)
cannot be a tensor'' (it is only a pseudo-tensor) and that
``it is not possible to obtain an expression for the energy of the
gravitational field satisfying both the conditions:
i)when added to other forms of energy the total energy is conserved,
and  ii)the energy within a definite (three dimensional) region at a
certain time is independent of the coordinate system.
Thus,
in general,
gravitational energy cannot be localized.
The best we can do is to use the pseudo-tensor which satisfies
condition i) but not ii).
It gives us approximate information about gravitational energy,
which in some special cases can be accurate''. Dirac dixit!
\citep{dir1975}.
Referring to the energy of Dirac's gravitational waves,
Dirac states that ``owing to the pseudo-tensor not being a real tensor,
we do not get,
in general,
a clear result independent of the coordinate system.
But there is one special case in which we do get a clear result;
namely,
when the waves are all moving in the same direction.'' \citep{dir1975}.
In the face of all this,
it is hard for us to see the need to cling to Einstein's General Theory
of Relativity with its unnecessary complicated non-linear field
equations in order to explain gravitational phenomena which can also be
explained from a linear system of vector equations such as (GEM),
which in addition predicts other phenomena in the astronomical scale.
Even such things as the hypothetical ``black holes'' could be
considered now within the framework of (GEM) with an analogue of strong
laser beams
(used for the source term $\vec{j}$)
employed by the 1997 Physics Nobel Prize winners Steven Chu,
Claude Cohen-Tannoudj\'i and William D. Phillips,
in an experiment aptly described as ``chilling with the light'',
whereby we see a piece of matter contracted almost to a point and
chilled near the absolute zero temperature.
In our vision of the interaction of light and matter,
this would be an earth lab scale analog of what could have been
``the big crunch'',
that is to say,
such an initial singularity might not be due to gravitational collapse
ensuing the end of the Universe expansion,
but it could be due to light pressure concentrated at a point in
$\hbox{E}_3$ due to highly energetic incoming light rays
``from infinity''.
Part of these light rays' energy would be reflected as outgoing light
rays in a discrete number of directions,
other part would be reflected as proto-matter,
which besides traveling away from 
``the explosion center point'',
would be propelled to rotate about each ray alongside due to the
gravitomagnetic field of the leading outgoing ray,
which would then be transferring its initial energy after reflection to
the rotating proto-nebulae as they move away,
and such nebulae would be kept approximately in planes normal to the
debilitating guiding reflected rays.
We shall come back to this view,
but in a mathematical way in section 4.

We should not end our historical review without setting the record
straight,
giving due credit to H. A. Lorentz,
who as far back as 1900,
that is to say far ahead of Wheeler and Mashhoon,
proposed the following, in his (originally written in Dutch) paper,
which later appeared in French
(``Considerations sur la pesanteur'' \cite{lor1937},
and from which we have freely translated to English,
also using modern notation for the electric and magnetic fields and
vector product,
keeping our rationalized system of units,

``It can be assumed that in the field of gravity there are two vectors
$\vec{E}$ and $\vec{B}$ determined by equations (I),
$\rho$ being the density of weighable matter,
and the force per unit mass shall be given by
$-\eta \left\{\vec{E} + \vec{v} \times \vec{B}\right\}$,
$\eta$ being a certain positive coefficient.

In every theory of gravitation it is important to examine which is the
influence of the motion of heavenly bodies upon their mutual action.
The answer to this question may be deduced from the preceding equations;
besides,
the problem is analogue in all points to a corresponding problem
relative to the electromagnetic actions between charged particles.
Etc., etc.''

Equations (I) referred to are given in page 208 of Lorentz' work (1937)
under the form (in modern notation):

\begin{eqnarray}
\vec{\nabla} \cdot \vec{D} &=& \rho
\nonumber \\
\nonumber \\
\vec{\nabla} \cdot \vec{B} &=& 0
\nonumber \\
\nonumber && \hskip 80pt (I) \\
\vec{\nabla} \times \vec{H} &=& \rho \vec{v} +
                             {\partial \vec{D} \over \partial t}
\nonumber \\
\nonumber \\
\vec{\nabla} \times \vec{E} &=& - {\partial \vec{B} \over \partial t}
\nonumber
\end{eqnarray}
 
\noindent
$\rho$ being the mass density at point $x$ and time $t$,
(recall that in Lorentz' electron theory,
 each particle with velocity $\vec{v}$ with respect to an ether
 assumed at rest,
 is likened to a small sphere with a continuous density
 distribution inside it and vanishing at its surface boundary).

From Ecclesiastes 1:9 ``The thing that hath been,
it is that which shall be,
and that which is done,
is that which shall be done:
and there is no new thing under the sun''.

Thus we see that Wheeler's intuition and Mashhoon gravitomagnetic
equations were long anticipated by Lorentz' genius.
Probably,
he didn't pursue this approach to gravitation because of his personal
acquaintance and friendship with Einstein,
whose towering figure by 1916 may have inhibited him from opposing
Einstein's views.

But more important to us is that Lorentz' theoretical formulation of
gravity as a strict analogue of electromagnetism,
as it had been developed by 1900,
in no way was tied to Einstein's theory of Gravitation,
which came up in 1916.

We follow his approach,
without reference to an underlying ether of course,
but besides a purely mass interaction,
we have postulated a version of his system of equations in order to
accommodate the force interaction between matter and light,
analog to Compton's procedure,
leading us a step beyond the simple mass-charge interchange with a sign
change in the source terms.
It is obvious that in Lorentz's formula the $\eta$ factor cannot
multiply at the same time the gravito-electric and the gravito-magnetic
force-field effects,
since in the limit of the stationary case we would get Newton's force
so that $\eta$ should be unity;
so that this factor will be taken to multiply only the gravito-magnetic
force term.
We propose to obtain its value in such a way that the gravito-magnetic
Larmor's precession formula yields the observed angular velocity of
precession of Mercury's orbit.
We have estimated the Sun's total angular momentum to be
$\approx 1.63 \times 10^{41} \; \hbox{kg} \, \hbox{m}^2 \,
 \hbox{s}^{-1}$ and have taken Mercury's orbital mean radius to be
$= 5.79 \times 10^{10} m$,
to get
$\omega_{Mercury} = 6.469 \times 10^{-12} \hbox{rad} \; \hbox{s}^{-1}$
with a dimensionless $\eta$ factor $= 4.152 \times 10^7$.

A final word about Quantum Gravity is appropriate.
Reading Hawking's speech at the 17th International Conference on
General Relativity and Gravitation,
Dublin,
given on July 21st 2004,
it doesn't seem that his Theory could be considered as an established,
definitive Gravitation Theory,
pending on the Maldacena conjecture,
and which further needs ``regularizing'' to cope with infinite action,
as Hawking puts it.
Thus, we feel that a theoretical space is opened for exact vector
gravitational waves as solutions of a familiar linear system of PDEs,
for which the modern quantum electrodynamics procedure,
already employed for Maxwell's equation \citep{fey1965},
can now be considered for GEM equations,
and this will be dealt with in Part II of the present work.
Perhaps some might think that taking distance from Einstein's
General Theory of Relativity would involve a denial of black holes.
We think that it is necessary to distinguish between physical
black holes,
if there are such,
and mathematical singularities
(Schwartzchild black hole, Kerr black hole, etc,),
which purportedly represent them,
along with their observable properties.
Different singularities may play the same role as those coming from
Einstein's and related gravitational field equations.
Our proposal includes singularities to be described in section 4,
and which we have baptized as ``Trio-Holes''.  

\section{The loss of energy in a binary star system due to
         gravitational waves.}

Since our vector gravitational wave equations are analogues to the
corresponding electromagnetic wave equations,
it is appropriate to recall some facts about electromagnetic radiation
emission.
In that case waves are formed by the time-change in the position and
distribution of the electric charges in the system,
and the same would be true when those ``charges'' are gravitational
masses instead of electric charges.

The simplest radiating systems are those represented by localized
oscillating systems of charge and current density,
as is done in classical electrodynamics where the analysis is carried
out on the basis of a multipole expansion \citep{jac1975}.
Remembering that the coefficients in that expansion are called
multipole moments,
and the total charge corresponds to the monopole moment,
and since changing the total amount of charge in the system with time
would violate the law of conservation of charge,
then the way to accomplish wave-radiation is to have a time varying
dipole moment,
in the case of the electromagnetic wave.
However in gravitational systems this is found to be impossible due to
the violation of the law of conservation of angular momentum,
so that one has to consider the next highest moment of mass
distribution,
the quadrupole moment,
for possible emission of gravitational radiation,
and this is coherent with the physics involved. 
The only known sources of gravitational waves strong enough to be
detected by any known means are astrophysical phenomena,
since only in deep space are massive bodies found moving fast enough to
produce a sizeable wave signal.
This phenomena includes binary star systems,
pulsars,
and many others.

Binary star systems are,
as their name implies,
composed of two stars.
The binary star system is one of the most common sources of
gravitational radiation and it is the simplest case understood.
Since our vector wave equations are formally the same as Maxwell's,
the loss of orbital energy when the stars orbit each other about their
center of gravity is explained mathematically in exactly the same way
as is done in the similar case of orbiting electric charges,
and this without invoking any General Relativity principles,
or invoking any approximation of low velocities compared to the
speed of light.
This illustrates the relative simplicity of the proposed gravitational
theory that conserves the principles of Special Relativity.

\section{The case of intense cosmogonic light beams and Trio-Holes}

We shall consider first an auxiliary quasi-steady photonic current
density defined by $\vec{j} = \vec{\alpha} \delta$ in an otherwise free
space,
with
$\delta = \delta(\vec{r})
        = \delta(x) \otimes \delta(y) \otimes \delta(z)$,
where $\vec{r} = x\hat{i} + y \hat{j} + z \hat{k}$  as usual,
$\vec{\alpha}$ is a directional vector and $\otimes$ stands for the
tensor product of one dimensional distributions,
amounting to a ``separation of variables procedure'' for handling
higher dimensional distributions when it is appropriate to do so. 

The associated auxiliary gravitomagnetic component $\vec{u}_1$ would
then,
according to equation (VGW2), satisfy:
$\Delta \vec{u}_1 = \vec{\nabla} \times \vec{j}$
and
$\vec{\nabla} \cdot \vec{u_1} = 0$,
with
$\vec{j} = \vec{\alpha} \delta$.

We are really interested in the gravitomagnetic field $\vec{u}_1^*$
generated by a very narrow semi-infinite light ray,
with a constant photonic current of intensity $I$,
either outgoing from the origin to infinity along the positive
$z$-axis direction,
or incoming from infinity in the reverse direction towards the origin,
where it is annihilated,
which is the case we shall study below.
The corresponding photonic current density $\vec{j^*}$  can then be
described in the following distributional form: 

\begin{equation}
\vec{j^*} = - I \hat{k} \delta(x) \otimes \delta(y) \otimes H(z)
\;\;\;\; .
\end{equation}

\noindent
where $H(z)$ is the Heaviside unit step function,
with the known property $H'(z)=\delta(z)$,
so that
$\vec{j} = -I \hat{k} \delta(x) \otimes \delta(y) \otimes \delta(z) =
 {\partial  \vec{j^*} / \partial z }$,
and correspondingly, $\vec{u}_1 ={\partial \vec{u}_1^* / \partial z}$.

Since $H(z) =1$ for $z > 0$,
$\int\int\vec{j^*} \cdot \hat{k} \, dx dy =
 -I \int\int \delta(x) \otimes \delta(y) \, dx dy = - I$
for $z>0$ when the domain of integration is any disk in a plane
orthogonal to the cosmogonic beam and centered at the origin.
However,
we want to find $\vec{u^*}_1$ without recourse to a hypothetical
analogue of Biot-Savart Law for differential elements of mass or
photonic currents,
which would be used,
in order to know beforehand the tangential character of $\vec{u}_1$
by means of its circulation as in electromagnetism.

For this purpose,
we shall profit from what we already know about Stokes' flows,
and the ``rotlet'' singular solutions. 
The Stokes' system for a viscous incompressible flow with unit
kinematic viscosity and density is: 

\begin{equation}
\Delta \vec{v} - \vec{\nabla} p = - \vec{b}
\;\;\;\; ,
\end{equation}

\noindent
where $\vec{b}$  is the body force,
and $\vec{\nabla} \cdot \vec{v} = 0$.
            .
The fundamental solution of this system,
associated to $\vec{b} = \vec{\alpha} \delta$,
has a pressure part,
$p = (\vec{\alpha} \cdot \vec{r}) / 4 \pi r^3$,
and a velocity part $S(\vec{\alpha})$.
Since $\vec{\nabla} \cdot  \vec{\nabla} \times \vec{v} =0$ identically,
it is readily found that,
after computing the Stokeslet $S(\vec{\alpha})$ of vector strength
$\vec{\alpha}$ given by

\begin{equation}
S(\vec{\alpha}) = {1 \over 8 \pi}
                  \left[ {\vec{\alpha} \over r} +
                        \left(\vec{\alpha} \cdot \vec{r} \right)
                        {\vec{r} \over r^3} \right]
\end{equation}

\noindent
[Fundamental solution of Ladyzhenskaya, see \citep{lad1964}],
then the ``rotlet'' solution defined by
$rotlet(\vec{\alpha})
= \vec{\nabla} \times \left(S(\vec\alpha)\right)$,
is such that
$\Delta(rotlet) = - \vec{\nabla} \times ( \vec{\alpha}\delta)$,
and explicitly one gets from its definition that 
$rotlet(\vec{\alpha}) = {2  \over r^3}(\vec{\alpha} \times \vec{r})$
(Power et~al. 1987),
and therefore we obtain that the auxiliary field
$\vec{u}_1 = rotlet(\vec{\alpha}) = {2 \over r^3}
 ( \vec{\alpha} \times \vec{r})$ 
is the gravitomagnetic field generated by a photonic current
density $\vec{j} = - \vec{\alpha} \delta$.

Finally,
since $\vec{u}_1 = \partial \vec{u^*}_1 / \partial z$,
we can obtain $\vec{u^*}_1$
by means of an integration with respect to the $z$-variable after
expressing $\vec{r}$ in cylindrical coordinates,
$\vec{r} = \rho \hat{\rho} + z \hat{k}$,
and $\vec{\alpha} = I \hat{k}$,
so that
$\vec{\alpha} \times \vec{r} = I \rho \hat{\theta}$,
$r^2 = \rho^2 + z^2$. 

Observe that for $z>0$, 

\begin{equation}
\vec{j^*}(x,y,z) =
\int\displaylimits_{-\infty}^z \vec{j}(x,y,t) \, dt =
- \int\displaylimits_{-\infty}^{z} I \hat{k} \delta(x) \otimes
                               \delta(y) \otimes \delta(t) \, dt =
- I \hat{k} \delta(x) \otimes \delta(y) \otimes H(z).
\end{equation}

The expression for $\vec{j^*}(x,y,z)$ is independent of ``$z$''
for $z>0$ by the above expression,
but that is not the case for $\vec{u^*}_1(x,y,z)$ as we show below:

\begin{equation}
u_1^*(x,y,z) - u_1^*(x,y,0) =
\int\displaylimits_0^z {\partial u^*_1 \over \partial z} dz =
\int\displaylimits_0^z
{2 I \rho \hat{\theta} \over (\rho^2 + z^2)^{3/2}} \, dz
=
{2 I \hat{\theta} \over \rho} \int\displaylimits_0^s
 (1 + s^2)^{-3/2} \, ds
=
{2 I \hat{\theta} \over \rho} {s \over (1+s^2)^{1/2} }
\;\;\;\; ,
\end{equation}

with $s = z / \rho$.
The $\hat{\theta}$ direction explains the rotation of
``proto-matter''
(this term we shall leave undefined for the time being)
about the $z$-axis.
As $z \longrightarrow +\infty$ with fixed $\rho$,
$\vec{u^*}_1(x,y,z)-\vec{u^*}_1(x,y,0) = 2 I \hat{\theta} / \rho$
asymptotically,
as one would expect,
since the semi-infinite photonic current tends to be seen,
for large $z>0$,
as an infinite filamentary current with a $1/\rho$ law for the
associated gravitomagnetic field $\vec{u}_1$ for all $z$.

The solution $\vec{u^*}_1$ thus obtained is only approximate,
being time independent,
since due to the photonic annihilation assumed to take place at the
origin,
and depending upon the physical nature of the process taking place
there
(for instance, a total conversion of light into matter which would
result at first from the simultaneous encounter at the origin of many
incoming light rays),
the mass density $\rho$ at that point would be increasing with time,
so that  $\rho(x_1,x_2,x_3,t) = m(t) \delta(x_1,x_2,x_3)$ and
$\vec{u}_2$ will vary with time,
with ${\partial \vec{u}_2 / \partial t}$ different from zero,
and now
$\vec{\nabla} \times \vec{u}_1 =
- \vec{j} + \epsilon_g \epsilon_{0g}
           (\partial \vec{u}_2 / \partial t)$
and
$\Delta \vec{u}_1 =
 - \vec{\nabla} \times \vec{j} + \epsilon_g \epsilon_{0g}
                  \vec{\nabla} \times
                  {\partial \vec{u}_2 / \partial t}$.
Since $\int\int\int \rho(x_1,x_2,x_3,t) \, dx_1 dx_2 dx_3 = m(t)$
measures the total mass accumulated at the singularity $(0,0,0)$
up to time ``$t$'',
the magnitude of the error made by neglecting
${\partial \vec{u}_2 / \partial t}$
will depend upon the time rate of change
$d m / dt$
of the increasing total mass $m(t)$ accumulated at $(0,0,0)$.

Now
$\vec{\nabla} \cdot \left( {\partial \vec{u}_2 / \partial t} \right)
 = - ( 1 / \epsilon_g \epsilon_{g0}) \cdot
     (\partial \rho / \partial t)
 = - ( 1 / \epsilon_g \epsilon_{g0}) \cdot (dm /dt) \cdot
     \delta(x_1,x_2,x_3)$
which has an obvious solution given by

\begin{equation}
{\partial \vec{u}_2 \over \partial t} =
\vec{\nabla} \left( {1 \over 4 \pi \epsilon_g \epsilon_{0g}} \cdot
{d m \over dt} \cdot {1 \over r} \right) =
{1 \over 4 \pi \epsilon_g \epsilon_{0g}} \cdot {d m \over dt}
\cdot \vec{\nabla}\left(1 \over r\right)
\end{equation}

\noindent
and

\begin{equation}
\vec{\nabla} \times \left( {\partial \vec{u}_2 \over \partial t} \right)
=
{1 \over 4 \pi \epsilon_g \epsilon_{0g}} \cdot {dm \over dt} \cdot
\vec{\nabla} \times \vec{\nabla} \left( 1 \over r \right) = 0
\;\;\;\; ,
\end{equation}

\noindent
so that it is still true that
$\Delta \vec{u}_1 = - \vec{\nabla} \times \vec{j}$,
but it will not be true that
$\vec{\nabla} \times \vec{u_1} = - \vec{j}$,
since now

\begin{equation}
\vec{\nabla} \times \vec{u}_1 =
- \vec{j} + {1 \over 4 \pi} \cdot {d \over dt} m \cdot
\vec{\nabla} \left( 1 \over r \right) 
\;\;\;\; .
\end{equation}

\noindent
 A better solution for the gravitomagnetic field
(under the hypothetical conditions modeling our incoming photonic
current) would be of the form 
$(\vec{u_1^*} + \vec{u_1^{**}})$,
where 

\begin{equation}
\vec{\nabla} \times \vec{u_1^{**}} =
{1 \over 4 \pi} \cdot {d \over dt}m \cdot \vec{\nabla}
\left(1 \over r \right)
\end{equation}

This result for $\vec{u_1^*}$,
even though approximate,
obviously can be applied to explain the angular momentum that can be
generated,
by both incoming and reflected outgoing strong light rays
(``kav'' lines of Luria)
along different $\vec{\alpha}$ directions,
in the cosmogonic processes after the Big Bang.
The angular directions can be quantized in a more detailed
Quantum Gravity model,
and we shall present it soon in part II of the present work.

It is also likely that this theory could have relevance in the atomic
scale, related to some spectral lines features.

\subsection{Trio-Holes}

In Einstein's geometrical approach to describe gravitational phenomena,
it has been customary to baptize ``Holes'' according to mathematical
singularities belonging to different metric tensors verifying the
non linear field equations,
and afterwards,
the Cosmos has been searched for astronomical evidence of their
``existence''.
Our approach goes in the opposite direction:
we make hypotheses about possible physical processes that take place
inside a given point,
and then look after an adequate mathematical description of it,
and then go to the Cosmos for corroboration. 

A singular point for our linear system of partial differential
equations may be the siege for three different processes,
namely:
light beam annihilated and completely transformed into matter,
light beam completely reflected,
and light beam partially reflected and partially converted into matter. 
This is the reason to speak of ``Trio-Holes'',
and for simplicity,
in this first paper,
we have worked out only the first process.
Also,
an exact time dependent solution for $\vec{u_1^*}$ can be obtained by
standard methods,
but for qualitative purposes only,
regarding the angular momentum present in the observed cosmos,
our approximate solution is sufficient. 

Besides,
the simplicity due to the linearity and familiarity of our proposed
vector wave equations,
allows the immediate treatment of complex situations by superposition
of solutions. 

\subsection{Universe mass previous to the Big-Bang}

We present a cosmological theory based in our view of gravitation
as a form of interaction between matter and light,
and in the equivalence of matter and energy,
according to Einstein's equation $E=mc^2$,
taken in both directions,
mass disintegration resulting into radiating energy,
and conversely,
matter resulting as a condensation of light,
and also based in the behavior of light photons in collisions with
matter as in the Compton effect.

We postulate that ``time'' begins with a separation of a pervading
light,
then occupying ``infinity'',
from an ``anti-light point'' which becomes a ``center''.
This ``polar'' situation results into photonic beams incoming from
``infinity'' into the center.
Initially,
all of these photons are totally converted into mass at the center.
As time goes by,
mass increases,
part of incoming light is reflected against the solid mass center,
and part continues converting into mass in a ``crunching'' process,
up to an instant when a critical magnitude is reached,
when the big explosion known as ``Big-Bang'' occurs.
Thus we can see matter in our present universe coming from different
origins and ages,
even before the ``Big-Bang'',
as some astronomical observations and measurements presently suggest.


\begin{thebibliography}{}

\bibitem[Alley 1995]{all1995b}
Alley, C. O. 1995, Annals of the New York Academy of Sciences,
Vol. 755 pp 464-475

\bibitem[Alley \& Yilmaz 1995]{all1995a}
Alley, C. O. \& Yilmaz, H. 1995, Science, Vol. 268, \#5210

\bibitem[Bleaney \&  Bleaney 1959]{ble1959}
Bleaney, B. I. \& Bleaney, B. 1959, Electricity and Magnetism
(Clarendon Press, Oxford)

\bibitem[De Broglie 1939]{deb1939}
De Broglie, L. 1939, Matter and Light, the New Physics,
1st English Edition
(from 1937 French by W. W. Norton \& Co. Inc., NY)

\bibitem[De Broglie 1952]{deb1952}
De Broglie, L. 1952, La F\'isica Nueva y los Cuantos
(Editorial Losada, Buenos Aires)

\bibitem[Dirac 1960]{dir1960}
Dirac P. A. M. 1960, Naturwissenschaftlishe Rundshau,
13 Jahrgang, pp 165-168

\bibitem[Dirac 1975]{dir1975}
Dirac, P. A. M. 1975,  General Theory of Relativity
(John Wiley \& Sons, Inc. NY)

\bibitem[Einstein 1918]{ein1918}
Einstein, A. 1918, Phys. Z., 19, pp 9-11

\bibitem[Einsten 1950]{ein1950}
Einstein, A. 1950, On the Generalized Theory, Scientific American

\bibitem[Feyman 1985]{fey1985}
Feynman, R. 1985, QED. The Strange Theory of Light and Matter
(Princeton University Press)

\bibitem[Feymann \& Hibbs 1965]{fey1965}
Feynman, R. P. \&  Hibbs,  A. R. 1965,
Quantum mechanics and Path Integrals
(McGraw-Hill Publishing Co., NY)

\bibitem[Gleick 1993]{gle1993}
Gleick, J. 1993, Genius: the Life and Science of Richard Feynman
(Vintage)

\bibitem[Hermann 1978]{her1978}
Hermann, R. 1978,  Yang-Mills,  Kaluza-Klein, and the Einstein Program
(Math Sci Press, Massachusetts)

\bibitem[Jackson 1975]{jac1975}
Jackson, J. D. 1975, Classical Electrodynamics 2nd Edition
(Wiley \& Sons, NY)

\bibitem[Ladyzhenskaya 1964]{lad1964}
Ladyzhenskaya, O. A. 1964, The Mathematical Theory of Viscous
Incompressible Flow 
(Gordon and Breach, NY)

\bibitem[Latapiat~2007]{lat2007}
Latapiat, R. 2007, A New Translation of Moses' Bereshit (Genesis) from
the Original Ruach Lashon Ha Kodesh
(to appear as Spanish and English new versions of the Bible,
supported by SOBICAIN.)

\bibitem[Lorentz 1909]{lor1909}
Lorentz, H. A. 1909, The Theory of Electrons
(Leiden reprinted by Dover)

\bibitem[Lorentz 1937]{lor1937}
Lorentz, H. A. 1937, Collected Papers, Vol. V
	(Martinus Nijhoff, The Hague)

\bibitem[Mashhoon 2003]{mas2003}
Mashhoon, B. 2003 Gravitoelectromagnetism: a Brief Review
              (http://www.arxiv.org/abs/gr-qc/0311030)

\bibitem[Miranda 1969]{mir1969}
Miranda, G. 1969, Ph.D. Thesis
(Purdue University, Supported by NASA Grant No. NGR 15-005-021)

\bibitem[Miranda \& Pereyra 2007]{mir2007a}
Miranda, G. \& Pereyra, N. 2007, Generalized Rankine-Hugoniot Shock
Condition and the Cauchy Problem for $u_y + u u_x = 0$ in the Presence
of a Moving Dirac Delta (unpublished)

\bibitem[Miranda \& Power 1983]{mir1983}
Miranda, G. \& Power, H. 1983,
Lecture Notes in Mathematics Vol. 1005, pp. 170-203
(Springer-Verlag, NY)

\bibitem[Miranda et al. 2007]{mir2007b}
Miranda, G., Power, H. \& Merentes, N. 2007,
The Space of Unbounded Harmonic Functions and the Dirichlet and Neumann
Problems in Unbounded Domains (unpublished).

\bibitem[Peplow 2004]{pep2004}
Peplow, M. 2004, Hawking Changes his Mind about Black Holes,
in news@nature.com, published online: July 15

\bibitem[Power \& Miranda 1987]{pow1987}
Power, H. \& Miranda G. 1987, Siam J. Appl. Math. Vol. 47, No 4
pp 689-698

\bibitem[Power \& Miranda 1986]{pow1986}
Power, H. \& Miranda G. 1986, On the Singular Solutions of Stokes and
Oseen's Equations, in Advances in Aerodynamics, Fluid Mechanics and
Hydraulics (ASCE)

\bibitem[Salam et~al. 1998]{sal1998}
Salam, A., Heisenberg, W., \& Dirac, P. 1998,
La Unificac\'ion de las Fuerzas Fundamentales (Gedisa)

\bibitem[Shr\"odinger 1918]{shr1918}
Schr\"odinger, E. 1918, Phys. Z., 19, pp 1-7

\bibitem[Schwartz 1959]{sch1959}
Schwartz, L. 1957, 1959, Theorie des Distributions, Tomes 1 \& 2
(Hermann, Paris)

\bibitem[Vladimirov 1984]{vla1984}
Vladimirov, V. S. 1984, Equations of Mathematical Physics
(Mir Publishers, Moscow)

\bibitem[Wheeler 1990]{whe1990}
Wheeler, J. A. 1990, A Journey into Gravity and Space Time
(Scientific American Library, NY)

\bibitem[Weyl 1915]{wey1915}
Weyl, H. 1915, Das asymptotische Verteilungsgesetz der
eigenschwingungen eines beliebig gestalteten elastischen K\"orpers,
Rend. d. Circ. Mat. di Palermo, Vol. 39, pp. 1-49 

\bibitem[Whittaker 1973]{whi1973}
Whittaker, E. 1973,  A History of the Theories of Aether and
Electricity
(Humanities Press, NY)

\bibitem[Zemanian 1987]{zem1987}
Zemanian, A. 1987, Distribution Theory and Transform Analysis
(Dover Publ. Inc., Mineola, NY)

\end{thebibliography}
\end{document}